\documentclass[aps,prl,twocolumn,
    showpacs,
    psuperscriptaddress,groupedaddress]{revtex4}
\usepackage{amssymb,amsmath,graphicx,color,microtype}

\begin{document}
\title{Towards the physical vacuum of cosmic inflation}

\author{Hongliang Jiang}
\author{Yi Wang}
\email{phyw@ust.hk}
\affiliation{Department of Physics, The Hong Kong University of Science and Technology,
Clear Water Bay, Kowloon, Hong Kong, P.R.China}


\begin{abstract}
  There have been long debates about the initial condition of inflationary perturbations. In this work we explicitly show the decay of excited states during inflation via interactions. For this purpose, we note that the folded shape non-Gaussianity can be interpreted as the decay of the non-Bunch-Davies initial condition. The one loop diagrams with non-Bunch-Davies propagators are calculated to uncover the decay of such excited states. The observed smallness of non-Gaussianity keeps the window open for probing inflationary initial conditions and trans-Planckian physics.
\end{abstract}

\maketitle

After decades of study of cosmic inflation, the initial condition for inflationary perturbations still remains a mystery. There is a long lasting debate on whether it is interesting to study non-trivial initial conditions, between the following points of view:
\begin{itemize}
  \item Inflation is an attractor solution. Thus all non-vacuum initial states should be washed away exponentially during inflation. So it is argued that only the lowest energy vacuum state is left, known as the Bunch-Davies (BD) vacuum \cite{Bunch:1978yq}.
  \item The non-vacuum states, once imposed, can stay for arbitrarily long time if the cosmological perturbations were linear. This is because, by definition, the linear perturbation theory relies on the initial condition in a trivial way. A modification of initial condition implies different integration constants of the inflaton equation of motion, and thus gets directly imprinted on the cosmic microwave background and the large scale structure \cite{Easther:2001fi}.
\end{itemize}

There are indeed many reasons to take the second viewpoint seriously. The non-vacuum initial states for the inflationary perturbations are supported by a number of considerations:

\begin{itemize}
  \item Just enough inflation. If inflation starts not long before the largest observable scales exit the horizon, the initial condition of those largest observable scales cannot be the inflationary BD vacuum. Because the BD vacuum is the lowest energy state of quasi-de Sitter background. Before inflation starts, the universe has a different background \cite{Chen:2013tna}.
  \item Features during inflation. The features in the inflaton Lagrangian, either one-off or periodic, kick the initial state of perturbations. The former excites all perturbation modes at an initial time, up to the energy scale defined by the sharpness of the feature, and the latter continuously excite perturbation modes \cite{Chen:2010bka}.
  \item Trans-$\Lambda$ problem. During the de Sitter expansion, the physical wavelength of perturbation modes starts off at small length scales and gets stretched. If inflation lasts long enough, when tracking back in time, the physical wavelength of any perturbation mode starts shorter than the UV cutoff $\Lambda$ of the IR effective field theory. As a result, new physics enters, such as new massive states or new operators. The most famous problem of such kind is the trans-Planckian problem \cite{Martin:2000xs}, because all known effective field theories have a universal cutoff $M_p$, above which quantum gravity effects kick in.
  \item The selection of non-linear vacuum. The practical definition of the inflationary vacuum is the lowest energy state for inflationary initial conditions. However, ambiguity of this definition arises when this practical approach is applied to gravitational fluctuations at nonlinear order. This is because the definition of energy, and thus the lowest energy state, depends on the choice of time coordinate, and thus is gauge dependent. The best known effect of such kind in empty Minkowski space is the Unruh radiation, that accelerated observer finds him/herself in a thermal state. At non-linear orders for cosmological perturbations, the curvature fluctuation $\zeta$ shifts the definition of time, and leaves ambiguity of the definition of vacuum in a similar way \cite{Chen:2012ye}.
\end{itemize}

In view of the above considerations, the non-trivial initial states are considered to be an interesting possibility, and can be parameterized by the non-BD coefficients $C_+(\mathbf{k})$ and $C_-(\mathbf{k})$, such that for any fluctuating field $\phi_\mathbf{k}$
\begin{equation}
  \phi_\mathbf{k} = u_\mathbf{k} a_\mathbf{k}
  + u^*_{-\mathbf{k}} a^\dagger_{-\mathbf{k}}~,
\end{equation}
where the mode functions are
\begin{equation}
  u_\mathbf{k} \equiv C_+(\mathbf{k}) u_k^\mathrm{BD}
  + C_-(\mathbf{k}) \left(u_k^\mathrm{BD}\right)^*~,
\end{equation}
where $k\equiv |\mathbf{k}|$  and $u_k^\mathrm{BD}$ is the Bunch-Davies vacuum, practically selected by the lowest energy state at early times. The consistency of uncertainty principle (commutation relation between $\phi_\mathbf{k}$ and its conjugate momentum) and the size of a quanta (commutation relation between $a_\mathbf{k}$ and $a^\dagger_{\mathbf{k}'}$) requires
\begin{equation}
  \left |  C_+(\mathbf{k}) \right |^2
  - \left |  C_-(\mathbf{k}) \right |^2
  = 1~.
\end{equation}
Satisfying the above condition, the non-BD coefficients are then largely unconstrained for current researches, given that their back-reaction to the background energy density remains small \cite{Chen:2006nt, Flauger:2013hra, Aravind:2013lra}. As a result, in the sub-horizon limit $|k\tau| \gg 1$, and in the weak non-BD limit $|C_-(\mathbf{k})| \ll 1$, the inflationary two point function for the curvature perturbation $\zeta$ gets a correction
\begin{equation}
  \langle \zeta_{\mathbf{k}}(\tau) \zeta_{\mathbf{k}'}(\tau) \rangle'
  = \langle\zeta_{\mathbf{k}}(\tau) \zeta_{\mathbf{k}'}(\tau) \rangle_\mathrm{BD}'
  (1 + \Delta^{\mathrm{tree}}_\mathrm{non-BD}(\tau))
\end{equation}
where prime denotes that $(2\pi)^3\delta^3(\mathbf{k} + \mathbf{k}')$ is stripped, and
\begin{equation}
  \Delta^{\mathrm{tree}}_\mathrm{non-BD} (\tau)
  = -2 c_\mathbf{k} \cos(2k\tau+\theta_\mathbf{k})~,
\end{equation}
where $C_-(\mathbf{k})$ is parameterized with real parameters $c_\mathbf{k}$ and $\theta_\mathbf{k}$ as
\begin{equation}
  C_-(\mathbf{k}) \equiv  c_\mathbf{k} e^{i \theta_\mathbf{k}}~.
\end{equation}

The arbitrariness of choosing those non-BD coefficients seems to contradict the intuition that inflation is an attractor solution and all non-ground states should be diluted.

In this letter, we clarify the relation between the possibility of non-BD initial condition and the attractor feature of inflation. The key observation is that, the decay of the non-BD states, if possible, must happen through interactions. The process is similar to the process towards thermal equilibrium, where short modes radiate long modes and the system finally stabilizes at the equilibrium state.

We start from commenting on the non-Gaussianities from the non-BD states. It is well known that the non-BD non-Gaussianities peaks at the folded limit \cite{Chen:2006nt, Chen:2009bc}. For $n$-point correlation function, the limit is $k_1 = \sum_{i=2}^n k_i$ (starting from 4-point correlation function and for large $C_-$, other folded configurations such as $k_1+k_2 = k_3+k_4$ also have poles). The physical interpretation is simple (but not noticed in the literature to the best of our knowledge). Once a perturbation mode is initially not in its lowest energy state, it is unstable and can decay into longer wavelength modes. Thus the folded limit of non-Gaussianity simply captures the decay of those non-BD coefficients.

\begin{figure}[htbp]
  \centering
  \includegraphics[width=0.3\textwidth]{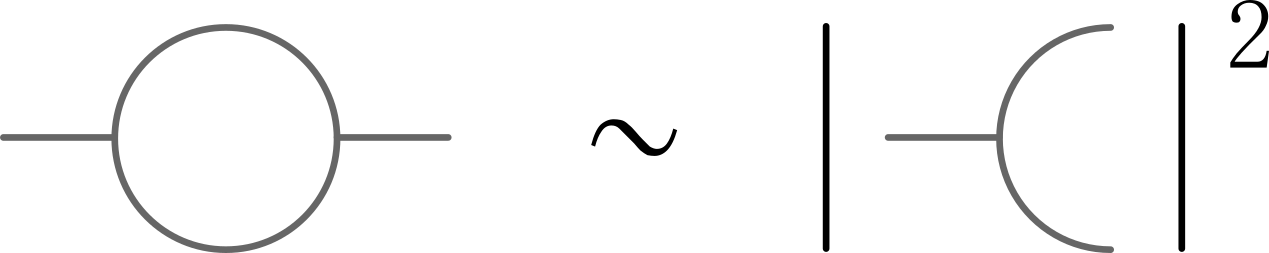}
  \caption{\label{fig:loop3} One-loop diagrams are needed to illustrate the decay of non-BD coefficients. And such one-loop diagrams are related to the three point function squared through the optical theorem.}
\end{figure}

To further clarify the situation, and make qualitative predictions, we would like to study explicitly how those nonlinear effects provide corrections to $C_+(\mathbf{k})$ and $C_-(\mathbf{k})$. For this purpose, we have to integrate over the decay products which we are not interested in, and consider the two point correlation function of the remaining high energy mode. This is nothing but the one loop diagram for the two point function, with non-BD coefficients. The relation between the non-Gaussianity and loop diagram is illustrated in Fig.~\ref{fig:loop3}.

Formally, the above argument is just the optical theorem, as a requirement of unitarity in QFT -- from probability conservation, the decay of non-BD states must be seen in both of the two ways: decay products appear, and the non-BD state itself decreases towards the BD vacuum, as
\begin{equation}\label{eq:decayRate}
  c^{\mathrm{eff}}_{\mathbf{k}}
  = c_{\mathbf{k}} \exp\left[ - \Gamma (\tau-\tau_0) \right]~,
\end{equation}
where $\tau_0$ is the initial time when the non-BD coefficients are generated, and $\Gamma$ is the decay rate (which may depend on $\mathbf{k}$ though not explicitly written here). $\Gamma$ can be time dependent, but the time dependence should be milder than linear on $\tau-\tau_0$.

In the remainder of the paper, we thus calculate the one loop correction to non-BD coefficients. To show the effect clearly, we choose the simplest possible model. Nevertheless, we expect that the decay of non-BD effect is completely general, and the method applies straightforwardly to other interactions.

The model we choose is general single field inflation, where $\mathcal{L} = \sqrt{-g} P(X, \phi)$. Further we require that the sound speed is still $c_s^2=1$, which is technically natural \cite{Baumann:2011su} from the effective field theory point of view, and provides us a simpler model for calculation. Namely, we require
\begin{equation}
  P_{,XX} = 0 ~, \quad \lambda \equiv X^2 P_{,XX} + \frac{2}{3} X^3 P_{,XXX} \neq 0.
\end{equation}

After those simplifications, the free Lagrangian for the curvature perturbation is
\begin{equation}
\mathcal{L}_2= \epsilon a^3 \dot{\zeta}^2-\epsilon a (\partial_i \zeta)^2 ~,
\end{equation}
and the 3rd order Lagrangian and Hamiltonian are
\begin{equation}
  \mathcal{L}_3 = -2a^3 \frac{\lambda}{H^3}\dot\zeta^3~,
  \qquad \mathcal{H}_3 = - \mathcal{L}_3 ~.
\end{equation}

Following the standard QFT quantisation procedure, the $\zeta$ field can be written as
\begin{equation}
\zeta_{\mathbf k}=u_{\mathbf k} a_{\mathbf k}+u_{-\mathbf k}^* a_{-\mathbf k}^\dagger ~,
\end{equation}
with the mode function
\begin{equation}
u_\mathbf{k}=\frac{H}{2 \sqrt{\epsilon k^3}}\Big[C_+( \mathbf k) (1+ik \tau) e^{-ik\tau}+C_-(\mathbf k )(1-i k\tau)e^{ik\tau}\Big]~.
\end{equation}

One can then calculate the one loop correction with non-BD coefficients making use of the in-in formalism. With two interaction vertices, as illustrated in the left panel of Fig.~\ref{fig:loop3}, the corrections can be calculated as
\begin{align} \label{eq:inin}
  &\langle\zeta_{\mathbf{k}}(\tau) \zeta_{\mathbf{k}'}(\tau) \rangle_3 =
  \int_{\tau_0}^\tau d\tau_1 \int_{\tau_0}^\tau d\tau_2
  \langle H_I(\tau_1)\zeta_{\mathbf{k}}(\tau) \zeta_{\mathbf{k}'}(\tau)H_I(\tau_2) \rangle
  \nonumber\\
  & - 2 \mathrm{Re}
  \int_{\tau_0}^\tau d\tau_1 \int_{\tau_0}^{\tau_1} d\tau_2
  \langle \zeta_{\mathbf{k}}(\tau) \zeta_{\mathbf{k}'}(\tau)H_I(\tau_1)H_I(\tau_2) \rangle ~,
\end{align}
where the subscript $3$ denotes that we are calculating the loop from $\mathcal{L}_3$.
The contribution to equation \eqref{eq:inin} breaks into three parts:
\begin{itemize}
  \item Part without any non-BD coefficients $C_-$, which is irrelevant for our purpose.
  \item Part proportional to $C_-(\mathbf{q})$ or $C_-(\mathbf{p})$, where $\mathbf{q}$ is the momentum running in the loop, and $\mathbf{p} \equiv \mathbf{k}-\mathbf{q}$. One should be able to find in this part that, once the loop contains a non-BD mode, how its energy is transferred to the other modes. This part of contribution does not directly show the decay of the non-BD mode. Mathematically, the contribution has an independent phase compared to $\Delta^{\mathrm{tree}}_\mathrm{non-BD}$. Thus we shall drop this part at the moment and leave it to a future work \cite{progress1}.
  \item Part proportional to $C_-(\mathbf{k})$. This shall be the part that we are going to calculate.
\end{itemize}
Further, we take the sub-horizon limit $|k\tau| \gg 1$. This is because at the super-horizon limit $|k\tau| \ll 1$, the perturbation is conserved and can no longer decay. The decay is dominated by the highest power of $|k\tau|$ terms, which are well captured in the $|k\tau| \gg 1$ limit.

Before showing the result, we would like to mention a subtlety here. Apparently, a divergence is encountered when we pick the highest order terms of $\tau$, by taking the $\tau\rightarrow -\infty$ limit. The momentum integral behaves as
\begin{equation}\label{eq:naive}
  \int d^3 q \frac{1}{p+q-k}~,
\end{equation}
Here the pole at $p+q-k \rightarrow 0$ arises because $\int d\tau \exp[i(p+q-k)\tau] \propto 1/(p+q-k)$. As a result, equation \eqref{eq:naive} diverges logarithmly. On the one hand, the pole at $1/(p+q-k)$ looks familiar, because it also appears at the calculation of non-Gaussianities, which shows evidence for the decay of non-BD. On the other hand, the contribution from this pole should not be physically infinity because otherwise it indicates an infinite decay rate, which is unphysical.

This problem can be resolved by noting that near the pole, $(p+q-k)(\tau-\tau_0)$ may be finite even when $k(\tau-\tau_0) \rightarrow \infty$. Thus we should not consider $(p+q-k)(\tau-\tau_0)$ to be always large in the $|k\tau| \gg 1$ expansion.

For this purpose, we split the momentum integral into $p+q-k<\Lambda$ and $p+q-k>\Lambda$ parts, where $\Lambda \simeq 1/(\tau-\tau_0)$. The $p+q-k<\Lambda$ part resolves the pole, where the contribution at $p+q-k$ is no longer singular. This $p+q-k<\Lambda$ contribution turns out to be a correction proportional to $\cos(\theta_{\mathbf{k}}+2k\tau)$, which corrects the amplitude $c_\mathbf{k}$ of the non-BD coefficient. On the other hand, the $p+q-k>\Lambda$ part, including the UV divergence (which should be renormalized), is proportional to $\sin(\theta_{\mathbf{k}}+2k\tau)$, which indicates a shift of phase $\theta_\mathbf{k}$, and has no contribution to $c_\mathbf{k}$ at the leading order of the small $c_\mathbf{k}$ expansion. We are mainly interested in the decay of the non-BD amplitude, thus shall currently focus on the $p+q-k<\Lambda$ part of the integral. We shall leave the full one-loop renormalized result in a future work \cite{progress1}.

After noting the above comments, the contribution near the apparent pole is straightforward to calculate. The leading contribution to the one loop two point function is
\begin{equation}
  \Delta_\mathrm{non-BD} = \Delta^{\mathrm{tree}}_\mathrm{non-BD}
  + \Delta^{\mathrm{1-loop}}_\mathrm{non-BD} + \cdots ~,
\end{equation}
where
\begin{eqnarray}\label{eq:res}
  &&
  \frac{\Delta^{\mathrm{1-loop}}_\mathrm{non-BD}}{\Delta^{\mathrm{tree}}_\mathrm{non-BD}}
  \equiv
  \frac{c_\mathbf{k}^\mathrm{eff}(\tau)-c_\mathbf{k}}{c_\mathbf{k}}
  = - \frac{1}{120} \left(\frac{\lambda}{\Sigma}\right)^2 P_\zeta k^5 \Lambda(\tau^3-\tau_0^3)^2   \nonumber \\
  &&\approx - \frac{1}{120} \left(\frac{\lambda}{\Sigma}\right)^2 P_\zeta k^5
  (\tau-\tau_0)(\tau^2 + \tau \tau_0 + \tau_0^2)^2~,
\end{eqnarray}
where we have chosen $\Lambda\approx 1/(\tau-\tau_0)$ although there may be a coefficient discrepancy.

It is interesting to note that this contribution is negative definite, indicating a decay of the non-BD coefficient. The full renormalization procedure may contribute to a change of coefficient in equation \eqref{eq:res}, but shall not change the sign or magnitude of the contribution.

When $k(\tau-\tau_0)$ is very large, the one-loop contribution can dominate over the tree level contribution and the result becomes non-perturbative. And this case is indeed physically interesting because we are interested to study the limit where the non-BD coefficient significantly decays. Fortunately, one can resum the one loop diagram to address the non-perturbative decay of the non-BD coefficients, by dividing $\tau-\tau_0$ into smaller intervals. Once the intervals are small enough, the contribution is guaranteed to be small because equation \eqref{eq:res} is proportional to the length of the interval. This process is formally nothing but the flow of renormalization group (RG). To do this, we rewrite equation \eqref{eq:res} into \footnote{Note that we still need $k(\tau-\tau_0) \geq 1$. Otherwise we are forcing interactions to take place at coincident time, and the other contributions will be missing.}
\begin{equation}
  d\log c_\mathbf{k}^\mathrm{eff}(\tau) \approx
  - \frac{3}{40} \left(\frac{\lambda}{\Sigma}\right)^2 P_\zeta k^5
  \tau^4 d\tau~~,
\end{equation}
where the integration constant should be determined to be $c_\mathbf{k}^\mathrm{eff}(\tau_0)=c_\mathbf{k}$. As a result, in terms of equation \eqref{eq:decayRate}, the decay rate is
\begin{equation}
  \Gamma = \frac{3}{200} \left(\frac{\lambda}{\Sigma}\right)^2 P_\zeta k^5
  (\tau^4 + \tau^3 \tau_0 + \tau^2\tau_0^2 + \tau \tau_0^3 + \tau_0^4)~.
\end{equation}

In terms of the non-Gaussianity estimator
$f_\mathrm{NL}^\lambda = - \frac{10}{81} \frac{\lambda}{\Sigma}$,
the conformal decay rate is
\begin{equation}
  \Gamma = \frac{19683}{20000} \left(f_\mathrm{NL}^\lambda\right)^2 P_\zeta k^5
  (\tau^4 + \tau^3 \tau_0 + \tau^2\tau_0^2 + \tau \tau_0^3 + \tau_0^4)~.
\end{equation}

Large non-Gaussianities have been pursued for two decades and a lot of people have been disappointed by the lack of large non-Gaussianity from recent experiments. However, here we show that large non-Gaussianity indicates faster decay of the non-trivial initial state of inflationary perturbations. This is the price to pay for large non-Gaussianity. Nature is equally kind to keep non-Gaussianity small, which leaves a broader window open for probing non-standard vacua, and thus gives us more hope for probing how inflation gets started, features during inflation, trans-Planckian effects, and so on.

For example, for $|f_\mathrm{NL}^\lambda| \sim 100$, which is of order the current observational bound, the non-BD $k$-modes which are 2 or more e-folds inside the horizon decay away significantly. For $|f_\mathrm{NL}^\lambda| \sim 1$, which is the minimal requirement for the inflationary background dynamics being nonlinear, significant decay occurs at 4 or more e-folds. For $|f_\mathrm{NL}^\lambda| \sim 0.01$, which is the slow roll bound given the detection of $n_s\neq 1$, the decay happens at 6 or more e-folds. The non-BD physics for modes with larger $k$ are exponentially harder to probe.

\begin{figure}[htbp]
  \centering
  \includegraphics[width=0.12\textwidth]{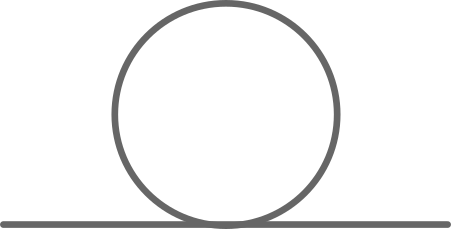}
  \caption{\label{fig:loop4} The 4-point contact interaction does not contribute to the decay of non-BD coefficients at one loop level.}
\end{figure}

For the purpose of comparison, we also calculated the one loop diagram from the four point contact interaction, with Lagrangian
\begin{equation}
  \mathcal{L}_4 = a^3 \frac{\mu}{H^4}\dot\zeta^4~,
\end{equation}
where $\mu$ can be obtained from the $P(X, \phi)$ model as
\begin{equation}
  \mu = \frac{1}{2} X^2 P_{,XX} + 2X^3 P_{,XXX} + \frac{2}{3} X^4 P_{,XXXX}~.
\end{equation}
The diagram is illustrated in Fig.~\ref{fig:loop4}. For this diagram, there is no $k$-dependent non-BD pole as the $\mathcal{L}_3$ case. This is because, physically, this diagram is not related to the three point function by the optical theorem. Mathematically, the external momentum $k$ does not enter the loop.

To summarize, we calculate the one loop correction to inflationary two point function, with non-BD coefficients. This calculation uncovers the decay of non-trivial initial states for inflationary perturbations. The conformal decay rate of the non-BD coefficients are proportional to $f_\mathrm{NL}^2 P_\zeta k^5 \tau^4$. Thus smaller $f_\mathrm{NL}$ indicates slower decay of the non-BD coefficients, and better preserves information from either beginning of inflation, features during inflation, or evidences of higher energy scales during inflation.

The calculation can inspire a lot of future work. Here we only consider the dominant contribution from the simplest model of interactions. A full calculation of the current model is in progress \cite{progress1}, and investigation of more general models are valuable. Also, it is interesting to go beyond the $|C_-| \ll |C_+|$ assumption, which is currently imposed in our calculation. The non-BD mode may also decay into more than two softer particles, which corresponds to two or more loop corrections to the two point function of the curvature fluctuation. We also plan to investigate explicitly how the folded limit non-Gaussianity get regularized because of the decay of the non-BD vacuum \cite{progress2}.

\noindent \textit{Acknowledgments.}
We thank Xingang Chen for his many valuable suggestions at the early stage of this work. We thank Gary Shiu for helpful discussions. YW and HJ are supported by a startup grant from the Hong Kong University of Science and Technology.

\end{document}